\documentclass[sigconf]{acmart}

\usepackage{booktabs} 
\usepackage[colorinlistoftodos]{todonotes}
\usepackage{balance}

\newcommand{\rel}{\textrm{rel}}
\newcommand{\rank}{\textrm{rank}}
\newcommand{\rr}{r}
\newcommand{\RR}{R}
\newcommand{\qq}{q}
\newcommand{\dd}{d}

\newcommand{\Prob}{\mathbf{P}}

\newcommand{\ubar}[1]{\textrm{U}(#1|\qq)}
\newcommand{\exposure}[1]{\textrm{Exposure}(#1|\Prob)}
\newcommand{\ctr}[1]{\textrm{CTR}(#1|\Prob)}

\newcommand{\Util}{\textrm{U}}
\newcommand{\bv}{\mathbf{v}}
\newcommand{\bu}{\mathbf{u}}
\newcommand{\D}{\mathcal{D}}
\newcommand{\user}{u}
\newcommand{\users}{\mathcal{U}}

 \usepackage[bb=boondox]{mathalfa}
 \usepackage{subcaption}
\usepackage[export]{adjustbox}
\usepackage{empheq}

\makeatletter
\renewcommand*{\p@section}{\S\,}
\renewcommand*{\p@subsection}{\S\,}
\renewcommand*{\p@subsubsection}{\S\,}
\makeatother


\DeclareMathOperator{\argmax}{argmax}

\DeclareMathOperator{\argsort}{argsort}
\usepackage{amsthm}

\copyrightyear{2018} 
\acmYear{2018} 
\setcopyright{acmlicensed}
\acmConference[KDD '18]{The 24th ACM SIGKDD International Conference on Knowledge Discovery \& Data Mining}{August 19--23, 2018}{London, United Kingdom}
\acmBooktitle{KDD '18: The 24th ACM SIGKDD International Conference on Knowledge Discovery \& Data Mining, August 19--23, 2018, London, United Kingdom}
\acmPrice{15.00}
\acmDOI{10.1145/3219819.3220088}
\acmISBN{978-1-4503-5552-0/18/08}
\title{Fairness of Exposure in Rankings}

\author{Ashudeep Singh}
\affiliation{%
  \institution{Cornell University}
  \city{Ithaca} 
  \state{NY} 
  \postcode{14853}
}
\email{ashudeep@cs.cornell.edu}

\author{Thorsten Joachims}
\affiliation{%
  \institution{Cornell University}
  \city{Ithaca} 
  \state{NY} 
  \postcode{14853}
}
\email{tj@cs.cornell.edu}

\begin{document}

\begin{abstract}
Rankings are ubiquitous in the online world today. As we have transitioned from finding books in libraries to ranking products, jobs, job applicants, opinions and potential romantic partners, there is a substantial precedent that ranking systems have a responsibility not only to their users but also to the items being ranked. To address these often conflicting responsibilities, we propose a conceptual and computational framework that allows the formulation of fairness constraints on rankings in terms of exposure allocation. As part of this framework, we develop efficient algorithms for finding rankings that maximize the utility for the user while provably satisfying a specifiable notion of fairness. Since fairness goals can be application specific, we show how a broad range of fairness constraints can be implemented using our framework, including forms of demographic parity, disparate treatment, and disparate impact constraints. We illustrate the effect of these constraints by providing empirical results on two ranking problems. 

\end{abstract}

%
%



\maketitle

\section{Introduction}
\label{section:1}
Rankings have become one of the dominant forms with which online systems present results to the user. Far surpassing their conception in library science as a tool for finding books in a library, the prevalence of rankings now ranges from search engines and online stores, to recommender systems and news feeds. Consequently, it is no longer just books that are being ranked, but there is hardly anything that is {\em not} being ranked today -- products, jobs, job seekers, opinions, potential romantic partners. Nevertheless, one of the guiding technical principles behind the optimization of ranking systems still dates back to four decades ago -- namely the Probability Ranking Principle (PRP)  \cite{robertson1977probability}. It states that the ideal ranking should order items in the decreasing order of their probability of relevance, since this is the ranking that maximizes utility  of the retrieval system to the user for a broad range of common utility measures in Information Retrieval. But is this uncompromising focus on utility to the users still appropriate when we are not ranking books in a library, but people, products and opinions?

There are now substantial arguments and precedent that many of the ranking systems in use today have responsibility not only to their users, but also to the items that are being ranked. In particular, the scarce resource that ranking systems allocate is the exposure of items to users, and exposure is largely determined by position in the ranking -- and so is a job applicant's chances to be interviewed by an employer, an AirBnB host's ability to rent out their property, or a writer to be read. This exposes companies operating with sensitive data to legal and reputation risks, and disagreements about a fair allocation of exposure have already led to high-profile legal challenges such as the European Union antitrust violation fine on Google \cite{googleeufine}, and it has sparked a policy debate about search neutrality \cite{grimmelmann2010some}. It is unlikely that there will be a universal definition of fairness that is appropriate across all applications, but we give three concrete examples where a ranking system may be perceived as unfair or biased in its treatment of the items that are being ranked, and where the ranking system may want to impose {\em fairness constraints} that guarantee some notion of fairness. 

The main contribution of this paper is a conceptual and computational framework for formulating fairness constraints on rankings, and the associated efficient algorithms for computing utility-maximi\-zing rankings subject to such fairness constraints. This framework provides a flexible way for balancing fairness to the items being ranked with the utility the rankings provide to the users. In this way, we are not limited to a single definition of fairness, since different application scenarios probably require different trade-offs between the rights of the items and what can be considered an acceptable loss in utility to the user. We show that a broad range of fairness constraints can be implemented in our framework, using its expressive power to link exposure, relevance, and impact. In particular, we show how to implement forms of demographic parity, disparate treatment, and disparate impact constraints. The ranking algorithm we develop provides provable guarantees for optimizing expected utility while obeying the specified notion of fairness in expectation.

To motivate the need and range of situations where one may want to trade-off utility for some notion of fairness, we start with presenting the following three application scenarios. They make use of the concept of protected groups\footnote{Groups that are protected from discrimination by law, based on sex, race, age, disability, color, creed, national origin, or religion. We use a broader meaning of protected groups here that suits our domain.}, where fairness is related to the differences in how groups are treated. However, we later discuss how this extends to individual fairness by considering groups of size one. The three examples illustrate how fairness can be related to a biased allocation of opportunity, misrepresentation of real-world distributions, and fairness as a freedom of speech principle.

\begin{figure}
    \centering
    \includegraphics[width=0.47\textwidth]{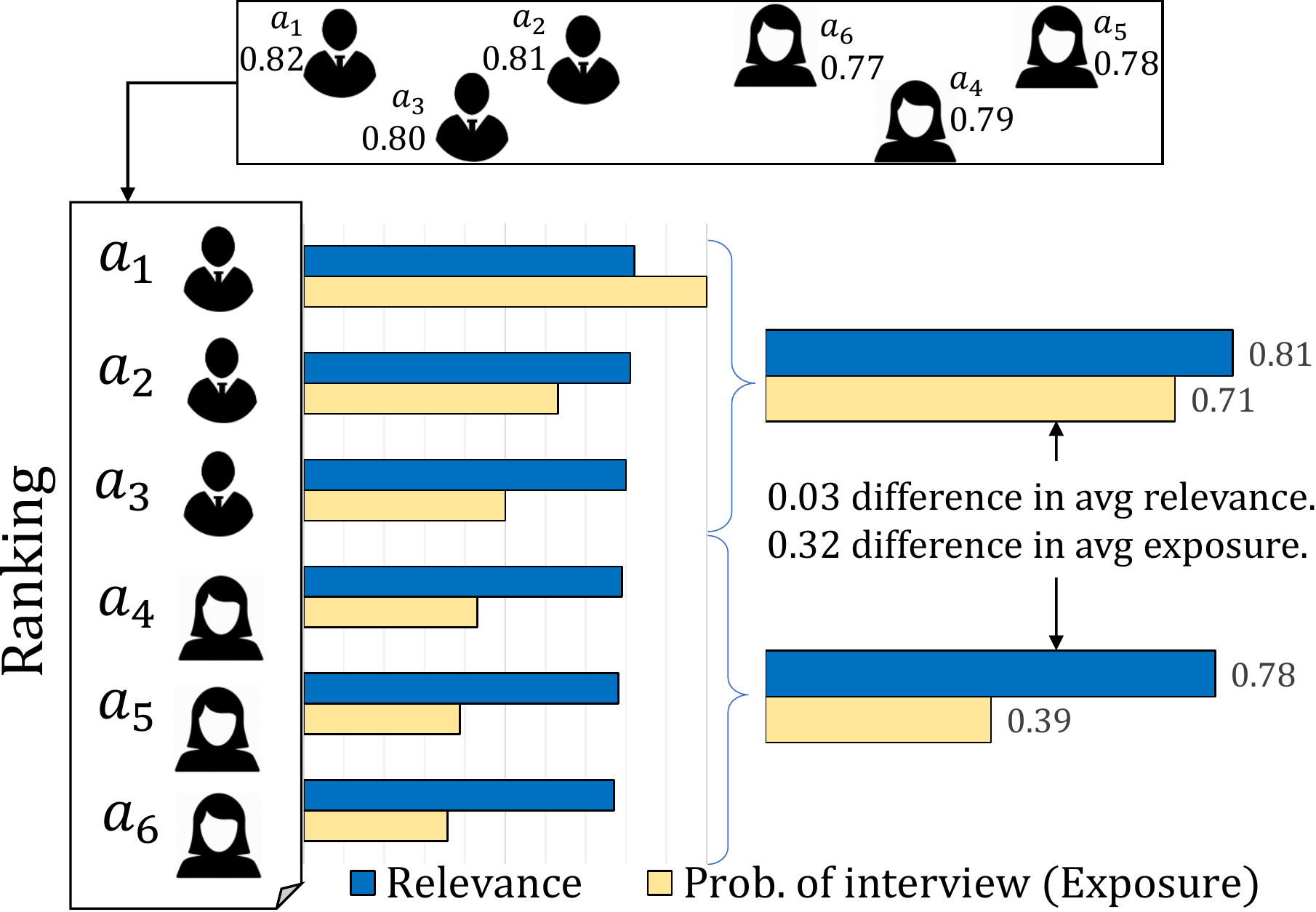}
    \caption{\textit{Job seeker example} to illustrate how small a difference in relevance can lead to a large difference in exposure (an opportunity) for the group of females.}
    \label{fig:toy-cartoon}
\end{figure}

\paragraph{Example 1: Fairly Allocating Economic Opportunity.}

Consider a web-service that connects employers (users) to potential employees (items). The following example demonstrates how small differences in item relevance can cause a large difference in exposure and therefore economic opportunity across groups. In this case, the web-service uses a ranking-based system to present a set of 6 applicants for a software engineering position to relevant employers (Figure~\ref{fig:toy-cartoon}). The set contains 3 males and 3 females. The male applicants have relevance of 0.80, 0.79, 0.78 respectively for the employers, while the female applicants have relevance of 0.77, 0.76, 0.75 respectively. Here we follow the standard probabilistic definition of relevance, where 0.77 means that 77\% of all employers issuing the query find that applicant relevant. The Probability Ranking Principle suggests ranking these applicants in the decreasing order of relevance i.e. the 3 males at the top positions, followed by the females. What does this mean for exposure between the two groups? If we consider a standard exposure drop-off (i.e., position bias) of $1/\log(1+j)$, where $j$ is the position in the ranking, as commonly used in the Discounted Cumulative Gain (DCG) measure, the female applicants will get 30\% less exposure -- even though the average difference in relevance between male and female applicants is just 0.03 (see Figure~\ref{fig:toy-cartoon}). Is this winner-take-all allocation of exposure fair in this context, even if the winner just has a tiny advantage in relevance? \footnote{Note that this tiny advantage may come from 3\% of the employers being gender biased, but this is not a problem we are addressing here.} It seems reasonable to distribute exposure more evenly, even if this may mean a small drop in utility to the employers.

\paragraph{Example 2: Fairly Representing a Distribution of Results.}

\begin{figure}
    \centering
    \includegraphics[width=0.47\textwidth]{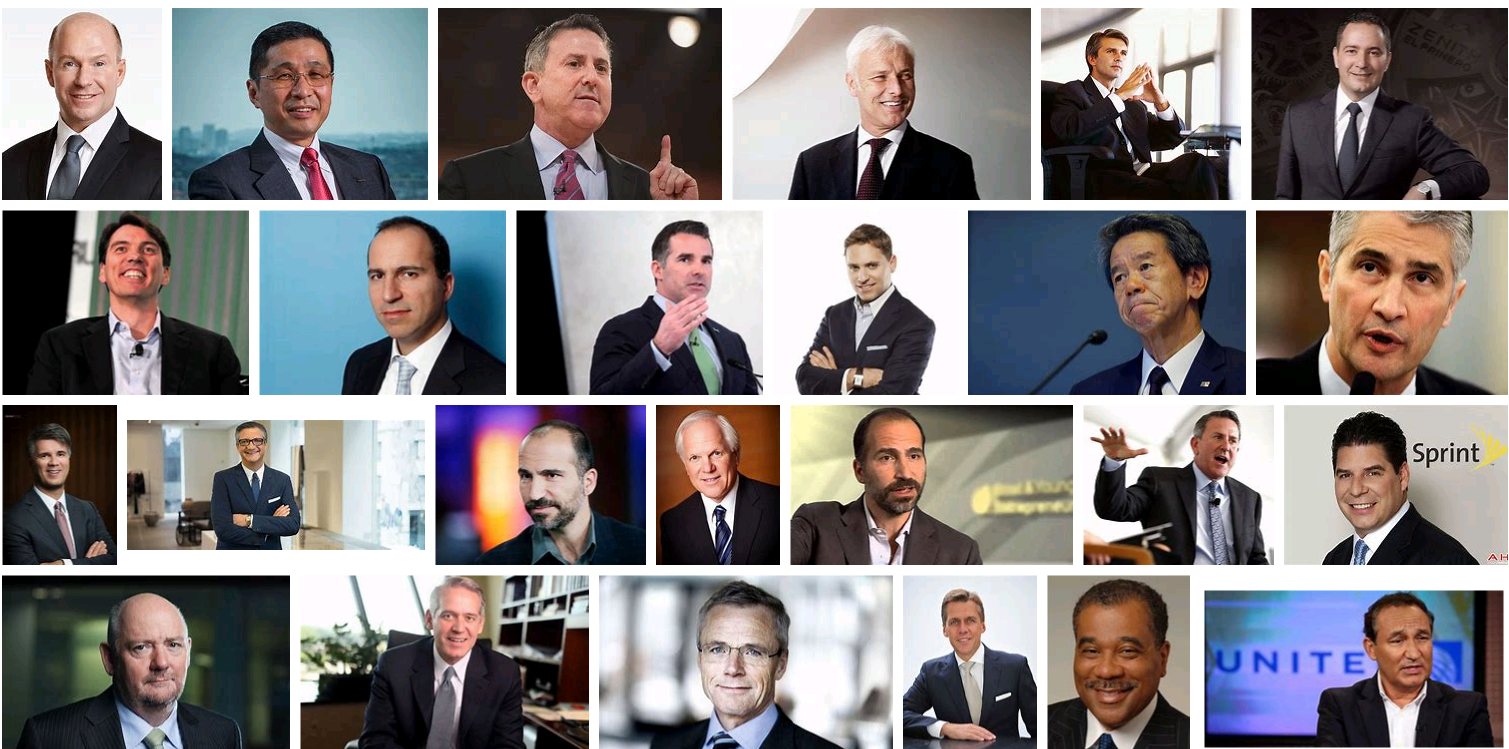}
    \caption{An image search result page for the query "CEO" showing a disproportionate number of male CEOs.}
    \label{img:barbie}
\end{figure}

Sometimes the results of a query are used as a statistical sample -- either explicitly or implicitly.  For example, a user may expect that an image search for the query ``CEO'' on a search engine returns roughly the right number of male and female executives, reflecting the true distribution of male vs.\ female CEOs in the world. If a search engine returns a highly disproportionate number of males as compared to females like in the hypothetical results in Figure~\ref{img:barbie}, then the search engine may be perceived as biased. In fact, a study detected the presence of gender bias in image search results for a variety of occupations \cite{Kay:2015:URG:2702123.2702520, bbc_barbie}. A biased information environment may affect users' perceptions and behaviors, and it was shown that such biases indeed affect people's belief about various occupations \cite{Kay:2015:URG:2702123.2702520}. Note that the Probability Ranking Principle does not necessarily produce results that represent the relevance distribution in an unbiased way. This means that even if users' relevance distribution agrees with the true distribution of female CEOs, the optimal ranking according to the Probability Ranking Principle may still look like that in Figure~\ref{img:barbie}. Instead of solely relying on the PRP, it seems reasonable to distribute exposure proportional to relevance, even if this may mean a drop in utility to the users. 


\paragraph{Example 3: Giving Speakers Fair Access to Willing Listeners.}
Ranking systems play an increasing role as a medium for speech, creating a connection between bias and fairness in rankings and principles behind freedom of speech \cite{grimmelmann2010some}. While the ability to produce speech and make this speech available on the internet has certainly created new opportunities to exercise freedom of speech for a speaker, there remains the question whether or not free speech makes its way to the interested listeners. Hence the study of the medium becomes necessary. Search engines are the most popular mediums of this kind and therefore have an immense capability of influencing user attention through their editorial policies, which has sparked a policy debate around search neutrality \cite{introna2000shaping, granka2010politics, grimmelmann2010some}. While no unified definition of search neutrality exists, many argue that search engines should have no editorial policies other than that their results are comprehensive, impartial, and solely ranked by relevance \cite{foundemnytimes}. But does ranking solely on relevance necessarily imply the Probability Ranking Principle, or are there other relevance-based ranking principles that lead to a medium with a more equitable distribution of exposure and access to willing listeners?

\section{Related Work}

Before introducing the algorithmic framework for formulating a broad range of fairness constraints on rankings, we first survey three related strands of prior work. First, this paper draws on concepts for algorithmic fairness of supervised learning in the presence of sensitive attributes. Second, we relate to prior work on algorithmic fairness for rankings. Finally, we contrast fairness with the well-studied area of diversified ranking in information retrieval.

\subsection{Algorithmic Fairness}
As algorithmic techniques, especially machine learning, find widespread applications, there is much interest in understanding its societal impacts. While algorithmic decisions can counteract existing biases, algorithmic and data-driven decision making affords new mechanisms for introducing unintended bias \cite{barocas2016big}. There have been numerous attempts to define notions of fairness in the supervised learning setting. The individual fairness perspective states that two individuals similar with respect to a task should be classified similarly \cite{dwork2012fairness}. Individual fairness is hard to define precisely because of the lack of agreement on task-specific similarity metrics for individuals. There is also a group fairness perspective for supervised learning that implies constraints like demographic parity and equalized odds. Demographic parity posits that decisions should be balanced around a sensitive attribute like gender or race \cite{calders2009building, zliobaite2015relation}. However, it has been shown that demographic parity causes a loss in the utility and infringes individual fairness \cite{dwork2012fairness}, since even a perfect predictor typically does not achieve demographic parity. Equalized odds represents the equal opportunity principle for supervised learning and defines the constraint that the false positive and true positive rates should be equal for different protected groups \cite{hardt2016equality}. Several recent works have focused on learning algorithms compatible with these definitions of fair classification \cite{zemel2013learning, woodworth2017learning, zafar2017fairness}, including causal approaches to fairness \cite{kilbertus2017avoiding, kusner2017counterfactual, nabi2017fair}. In this paper, we draw on many of the concepts introduced in the context of fair supervised learning but do not consider the problem of learning. Instead, we ask how to fairly allocate exposure in rankings based on relevance, independent of how these relevances may be estimated.

\subsection{Fairness in Rankings}
Several recent works have raised the question of group fairness in rankings. Yang and Stoyanovich \cite{yang2016measuring} propose statistical parity based measures that compute the difference in the distribution of different groups for different prefixes of the ranking (top-10, top-20 and so on). The differences are then averaged for these prefixes using a discounted weighting (like in DCG). This measure is then used as a regularization term. Zehlike et al. \cite{zehlike2017fa} formulate the problem of finding a `Fair Top-k ranking' that optimizes utility while satisfying two sets of constraints: first, in-group monotonicity for utility (i.e. more relevant items above less relevant within the group), second, a fairness constraint that the proportion of protected group items in every prefix of the \textit{top-k} ranking is above a minimum threshold. Celis et al. \cite{celis2017ranking} propose a constrained maximum weight matching algorithm for ranking a set of items efficiently under a fairness constraint indicating the maximum number of items with each sensitive attribute allowed in the top positions.
Some recent approaches, like Asudeh et al. \cite{asudehy2017designing}, have also looked at the task of designing fair scoring functions that satisfy desirable fairness constraints. 

Most of the fairness constraints defined in the previous work reflect parity constraints restricting the fraction of items with each attribute in the ranking \cite{Singh/Joachims/17a}. The framework we propose goes beyond such parity constraints, as we propose a general algorithmic framework for efficiently computing optimal probabilistic rankings for a large class of possible fairness constraints.

Concurrent and independent work by Biega et al. \cite{biega2018equity} formulates fairness for rankings similar to the special case of our framework discussed in Section~\ref{section:fairness2}, aiming to achieve amortized fairness of attention by making exposure proportional to relevance. They focus on individual fairness, which in our framework amounts to the special case of protected groups of size one. The two approaches not only differ in expressive power, but algorithmically, they solve an integer linear program to generate a series of rankings, while our approach provides a provably efficient solution via a standard linear program and the Birkhoff-von Neumann decomposition \cite{birkhoff1940lattice}.


\subsection{Information Diversity in Retrieval}
At first glance, fairness and diversity in rankings may appear related, since they both lead to more diverse rankings. However, their motivation and mechanisms are fundamentally different. Like the PRP, diversified ranking is entirely beholden to maximizing utility to the user, while our approach to fairness balances the needs of users and items. In particular, both the PRP and diversified ranking maximize utility for the user alone, their difference lies merely in the utility measure that is maximized. Under extrinsic diversity \cite{radlinski2009redundancy}, the utility measure accounts for uncertainty and diminishing returns from multiple relevant results \cite{Carbonell:1998:UMD:290941.291025,radlinski2008learning}. Under intrinsic diversity \cite{radlinski2009redundancy}, the utility measure considers rankings as portfolios and reflects redundancy \cite{Clarke:2008:NDI:1390334.1390446}. And under exploration diversity \cite{radlinski2009redundancy}, the aim is to maximize utility to the user in the long term through more effective learning. The work on fairness in this paper is fundamentally different in its motivation and mechanism, as it does not modify the utility measure for the user but instead introduces rights of the items that are being ranked.


\section{A Framework for Ranking under Fairness Constraints}
\label{section:3}

Acknowledging the ubiquity of rankings across applications, we conjecture that there is no single definition of what constitutes a fair ranking, but that fairness depends on context and application. In particular, we will see below that different notions of fairness imply different trade-offs in utility, which may be acceptable in one situation but not in the other. To address this range of possible fairness constraints, this section develops a framework for formulating fairness constraints on rankings, and then computing the utility-maximizing ranking subject to these fairness constraints with provable guarantees. 

For simplicity, consider a single query $\qq$ and assume that we want to present a ranking $\rr$ of a set of documents $\D = \{\dd_1, \dd_2, \dd_3\allowbreak \dots, \dd_N\}$. Denoting the utility of a ranking $\rr$ for query $\qq$ with $\Util(\rr|\qq)$, the problem of optimal ranking under fairness constraints can be formulated as the following optimization problem:
\begin{eqnarray*}
\rr & = &  \argmax_{\rr} \Util(\rr|\qq)\\
    &  & \mbox{s.t. $\rr$ is fair} 
\end{eqnarray*}
In this way, we generalize the goal of the Probabilistic Ranking Principle, which emerges as the special case of no fairness constraints. To fully instantiate and solve this optimization problem, we will specify the following four components. First, we define a general class of utility measures $\Util(\rr|\qq)$ that contains many commonly used ranking metrics. Second, we address the problem of how to optimize over rankings, which are discrete combinatorial objects, by extending the class of rankings to probabilistic rankings. Third, we reformulate the optimization problem as an efficiently solvable linear program, which implies a convenient yet expressive language for formulating fairness constraints. And, finally, we show how a probabilistic ranking can be efficiently recovered from the solution of the linear program.

\subsection{Utility of a Ranking}

Virtually all utility measures used for ranking evaluation derive the utility of the ranking from the relevance of the individual items being ranked. For each user $\user$ and query $\qq$, $\rel(\dd|\user,\qq)$ denotes the binary relevance of the document $d$, i.e. whether the document is relevant to user $\user$ or not. Note that different users can have different $\rel(\dd|\user,\qq)$ even if they share the same $\qq$. To account for personalization, we assume that the query $\qq$ also contains any personalization features and that $\users$ is the set of all users that lead to identical $\qq$. Beyond binary relevance, $\rel$ could also represent other relevance rating systems such as a Likert scale in movie ratings, or a real-valued score. 

A generic way to express many utility measures commonly used in information retrieval is
\[
    \Util(\rr|\qq) = \sum_{\user\in \users}P(\user|\qq)\sum_{\dd\in\D} v(\rank(\dd|\rr)) \lambda(\rel(\dd|\user,\qq)), 
\]
where $v$ and $\lambda$ are two application-dependent functions. The function $v(\rank(\dd|\rr))$ models how much attention document $d$ gets at rank $\rank(\dd|\rr)$, and $\lambda$ is a function that maps the relevance of the document for a user to its utility. In particular, the choice of $v$ could be based on the position bias i.e. the fraction of users who examine the document shown at a particular position out of the total number of users who issue the query $\qq$. The choice of $\lambda$ mapping relevance to utility is somewhat arbitrary. For example, a widely used evaluation measure, Discounted Cumulative Gain (DCG) \cite{jarvelin2002cumulated}
can be represented in our framework where $v(\rank(\dd|\rr))=\frac{1}{\log(1+\rank(\dd|\rr))}$, and $\lambda(\rel(\dd|\user,\qq))=2^{\rel(\dd|\user,\qq)}-1$ (or sometimes simply $\rel(\dd|\user,\qq)$): 
\[DCG(\rr|\qq) = \sum_{\user\in \users} P(\user|\qq) \sum_{\dd\in\D} \frac{2^{\rel(\dd|\user,\qq)}-1}{log(1+\rank(\dd|\rr))}\]
For a measure like $DCG@k(\rr|\qq)$, we can choose $v(\rank(\dd|\rr))= \frac{1}{\log(1+\rank(\dd|\rr))}$ for $\rank(\dd|\rr)\leq k$ and $v(\rank(\dd|\rr))=0$ for $\rank(\dd|\rr)>k$.\\

Since utility is linear in both $v$ and $\lambda$, we can combine the individual utilities into an expectation
\begin{align}
    \Util(\rr|\qq) &= \sum_{\dd\in\D} v(\rank(\dd|\rr)) \bigg(\sum_{\user\in \users} \lambda(rel(\dd|\user,\qq)) \: P(\user|\qq)\bigg) \notag{}\\
            &= \sum_{\dd\in\D} v(\rank(\dd|\rr)) u(\dd|\qq) \notag{}, 
\end{align}
where 
\[
u(\dd|\qq) = \sum_{\user\in \users} \lambda(rel(\dd|\user,\qq)) \: P(\user|\qq)
\]
is the expected utility of a document $\dd$ for query $\qq$. In the case of binary relevances and $\lambda$ as the identity function, $u(\dd|\qq)$ is equivalent to the probability of relevance. It is easy to see that sorting the documents by $u(\dd|\qq)$ leads to the ranking that maximizes the utility 
\[
\argmax_{\rr} \Util(\rr|\qq) \equiv  \argsort_{\dd\in\D} u(\dd|\qq)
\]
for any function $v$ that decreases with rank. This is the insight behind the Probability Ranking Principle (PRP) \cite{robertson1977probability}.


\subsection{Probabilistic Rankings} 

Rankings are combinatorial objects, such that naively searching the space of all rankings for a utility-maximizing ranking under fairness constraints would take time that is exponential in $|\D|$. To avoid such combinatorial optimization, we consider probabilistic rankings $\RR$ instead of a single deterministic ranking $\rr$. A probabilistic ranking $\RR$ is a distribution over rankings, and we can naturally extend the definition of utility to probabilistic rankings.
\begin{eqnarray*}
    \Util(\RR|\qq) &=& \sum_{\rr} \RR(\rr) \sum_{\user\in \users}P(\user|\qq)\sum_{\dd\in\D} v(\rank(\dd|\rr)) \lambda(\rel(\dd|\user,\qq)) \\
    & = &\sum_{\rr} \RR(\rr) \sum_{\dd\in\D} v(\rank(\dd|\rr)) \: u(\dd|\qq)
\end{eqnarray*}
While distributions $\RR$ over rankings are still exponential in size, we can make use of the additional insight that utility can already be computed from the marginal rank distributions of the documents. Let $\Prob_{i,j}$ be the probability that $\RR$ places document $d_i$ at rank $j$, then $\Prob$ forms a doubly stochastic matrix of size $N\times N$, which means that the sum of each row and each column of the matrix is equal to 1. In other words, the sum of probabilities for each position is 1 and the sum of probabilities for each document is 1, i.e. $\sum_i \Prob_{i,j} = 1$ and $\sum_j \Prob_{i,j} = 1$. 
With knowledge of the doubly stochastic matrix $\Prob$, expected utility for a probabilistic ranking can be computed as
\begin{align}
 \Util(\Prob|\qq) = \sum_{d_i\in\D}\sum_{j=1}^N \Prob_{i,j} \: u(d_i|\qq) \: v(j)  .
\end{align}
To make notation more concise, we can rewrite the utility of the ranking as a matrix product. For this, we introduce two vectors: $\mathbf{u}$ is a column vector of size $N$ with $\mathbf{u}_i = u(d_i|q)$ , and $\mathbf{v}$ is another column vector of size $N$ with $\mathbf{v}_j = v(j)$. So, the expected utility (e.g. DCG) can be written as:
\begin{align}
    \Util(\Prob|\qq) = \mathbf{u}^T \Prob \bv
\end{align}

\subsection{Optimizing Fair Rankings via Linear Programming} \label{optimization_prob}

We will see in Section~\ref{sec:bvn} that not only does $\RR$ imply a doubly stochastic matrix $\Prob$, but that we can also efficiently compute a probabilistic ranking $\RR$ for every doubly stochastic matrix $\Prob$. We can, therefore, formulate the problem of finding the utility-maximizing probabilistic ranking under fairness constraints in terms of doubly stochastic matrices instead of distributions over rankings. 
\begin{align}
\Prob = \argmax_{\Prob} \:\:&\mathbf{u}^T\Prob \mathbf{v} \tag{expected utility}\label{optimization}\\
    \textrm{s.t. } \:&\mathbb{1}^T \Prob = \mathbb{1}^T \tag{sum of probabilities for each position}\\
                        &\Prob \mathbb{1} = \mathbb{1} \tag{sum of probabilities for each document}\\
                        & 0 \leq \Prob_{i,j} \leq 1 \tag{valid probability}\\
                        &\Prob \mbox{ is fair} \tag{fairness constraints}
\end{align}
Note that the optimization objective is linear in $N^2$ variables $\Prob_{i,j}, 1\leq i, j \leq N$. Furthermore, the constraints ensuring that $\Prob$ is doubly stochastic are linear as well, where $\mathbb{1}$ is the column vector of size $N$ containing all ones. Without the fairness constraint and for any $\bv_j$ that decreases with $j$, the solution is the permutation matrix that ranks the set of documents in decreasing order of utility (conforming to the PRP). 

Now that we have expressed the problem of finding the utility-maximizing probabilistic ranking, besides the fairness constraint, as a linear program, a convenient language to express fairness constraints would be linear constraints of the form
\[
    \textbf{f}^T\Prob \textbf{g}=h. \notag{}
\]
One or more of such constraints can be added, and the resulting linear program can still be solved efficiently and optimally with standard algorithms like interior point methods. As we will show in Section~\ref{section:constraints}, the vectors $\mathbf{f}$, $\mathbf{g}$ and the scalar $h$ can be chosen to implement a range of different fairness constraints. To give some intuition, the vector $\mathbf{f}$ can be used to encode group identity and/or relevance of each document, while $\mathbf{g}$ will typically reflect the importance of each position (e.g. position bias). 

\subsection{Sampling Rankings} \label{sec:bvn}

The solution $\Prob$ of the linear program is a matrix containing probabilities of each document at each position. To implement this solution in a ranking system, we need to compute a probabilistic ranking $\RR$ that corresponds to $\Prob$. From this probabilistic ranking, we can then sample rankings $\rr \sim \RR$ to present to the user\footnote{For usability reasons, it is preferable to make this sampling pseudo-random based on a hash of the user's identity, so that the same user receives the same ranking $\rr$ if the same query is repeated.}. Given the derivation of our approach, it is immediately apparent that the rankings $\rr$ sampled from $\RR$ fulfill the specified fairness constraints in expectation.

Computing $\RR$ from $\Prob$ can be achieved via the Birkhoff-von Neumann (BvN) decomposition \cite{birkhoff1940lattice}, which provides a transformation to decompose a doubly stochastic matrix into a convex sum of permutation matrices. In particular, if $\mathbf{A}$ is a doubly stochastic matrix, there exists a decomposition of the form 
\[
\mathbf{A} = \theta_1 \mathbf{A_1} + \theta_2 \mathbf{A_2} + \dots + \theta_n \mathbf{A_n}
\]
where $0\leq \theta_i\leq 1$, $\sum_i \theta_i =1$, and where the $\mathbf{A_i}$ are permutation matrices \cite{birkhoff1940lattice}. In our case, the permutation matrices correspond to deterministic rankings of the document set and the coefficients correspond to the probability of sampling each ranking. According to the Marcus-Ree theorem, there exists a decomposition with no more than $(N-1)^{2}+1$ permutation matrices \cite{marcus1959diagonals}. Such a decomposition can be computed efficiently in polynomial time using several algorithms \cite{chang1999service, dufosse2016notes}. For the experiments in this paper, we use the implementation provided at \url{https://github.com/jfinkels/birkhoff}.

\subsection{Summary of Algorithm}

The following summarizes the algorithm for optimal ranking under fairness constraints. Note that we have assumed knowledge of the true relevances $u(\dd|\qq)$ throughout this paper, whereas in practice one would work with estimates $\hat{u}(\dd|\qq)$ from some predictive model.
\begin{enumerate}
\item Set up the utility vector $\mathbf{u}$, the position discount vector $\mathbf{v}$, as well as the vectors $\mathbf{f}$ and $\mathbf{g}$, and the scalar $h$ for the fairness constraints (see Section~\ref{section:constraints}).
\item Solve the linear program from Section~\ref{optimization_prob} for $\Prob$.
\item Compute the Birkhoff-von Neumann decomposition $\Prob = \theta_1 \Prob_1 + \theta_2 \Prob_2 + \dots + \theta_n \Prob_n$.
\item Sample permutation matrix $\Prob_i$ with probability proportional to $\theta_i$ and display the corresponding ranking $\rr_i$.
\end{enumerate}
Note that the rankings sampled in the last step of the algorithm fulfill the fairness constraints in expectation, while at the same time they maximize expected utility.

\section{Constructing Group Fairness Constraints}
\label{section:constraints}

Now that we have established a framework for formulating fairness constraints and optimally solving the associated ranking problem, we still need to understand the expressiveness of constraints of the form $\textbf{f}^T\Prob \textbf{g}=h$. In this section, we explore how three concepts from algorithmic fairness -- demographic parity, disparate treatment, and disparate impact -- can be implemented in our framework and thus be enforced efficiently and with provable guarantees. They all aim to fairly allocate exposure, which we now define formally. Let $\bv_j$  represent the importance of position $j$, or more concretely the position bias at $j$, which is the fraction of users that examine the item at this position. Then we define exposure for a document $\dd_i$ under a probabilistic ranking $\Prob$ as
\begin{eqnarray}
\exposure{\dd_i} = \sum_{j=1}^N \Prob_{i,j} \bv_j \label{exposure}
\end{eqnarray}

The goal is to allocate exposure fairly between groups $G_k$. Documents and items may belong to different groups because of some sensitive attributes -- for example, news stories belong to different sources, products belong to different manufacturers, applicants belong to different genders. The fairness constraints we will formulate in the following implement different goals for allocating exposure between groups. 

To illustrate the effect of the fairness constraints, we will provide empirical results on two ranking problems. For both, we use the average relevance of each document (normalized between 0 and 1) as the utility $\bu_i=u(\dd_i|\qq)$ and set the position bias to $\bv_j = \frac{1}{\log(1+j)}$ just like in the standard definition of DCG. More generally, one can also plug in the actual position-bias value, which can be estimated through an intervention experiment \cite{joachims2017unbiased}. 

\paragraph{Job-seeker example.} We come back to the job-seeker example from the introduction, and as illustrated in Figure~\ref{fig:toy-cartoon}. The ranking problem consists of 6 applicants with probabilities of relevance to an employer of $\bu=(0.81, 0.80, 0.79, 0.78, 0.77, 0.76)^T$. Groups $G_0$ and $G_1$ reflect gender, with the first three applicants belonging to the male group and the last three to the female group. 

\paragraph{News recommendation dataset.} We use a subset the \textit{Yow} news recommendation dataset \cite{zhang:2005} to analyze our method on a larger and real-world relevance distribution. The dataset contains explicit and implicit feedback from a set of users for news articles from different RSS feeds. We randomly sample a subset of news articles in the ``people'' topic coming from the top two sources. The sources are identified using RSS Feed identifier and used as groups $G_0$ and $G_1$. The `relevant' field is used as the measure of relevance for our task. Since the relevance is given as a rating from 1 to 5, we divide it by 5 and add a small amount of Gaussian noise ($\epsilon=0.05$) to break ties. The resulting $\bu_i$ are clipped to lie between 0 and 1. 

In the following, we formulate fairness constraints using three ideas for allocation of exposure to different groups. In particular, we will define constraints of the form $\textbf{f}^T \Prob \textbf{g} = h$ for the optimization problem in \ref{optimization_prob}. For simplicity, we will only present the case of a binary valued sensitive attribute i.e. two groups $G_0$ and $G_1$. However, these constraints may be defined for each pair of groups and for each sensitive attribute, and be included in the linear program.

\subsection{Demographic Parity Constraints} \label{section:fairness1}

Arguably the simplest way of defining fairness of exposure between groups is to enforce that the average exposure of the documents in both the groups is equal. Denoting average exposure in a group with
\[
\exposure{G_k}=\frac{1}{|G_k|}\sum_{d_i \in G_k} \exposure{d_i},
\]
this can be expressed as the following constraint in our framework:
\begin{align}
    &\:\:\exposure{G_0} = \exposure{G_1}\\
    \Leftrightarrow &\:\:\frac{1}{|G_0|}\sum_{d_i \in G_0} \sum_{j=1}^N \Prob_{i,j} \bv_j = \frac{1}{|G_1|}\sum_{d_i \in G_1} \sum_{j=1}^N \Prob_{i,j} \bv_j \\
    \Leftrightarrow & \:\:\sum_{d_i \in \D} \sum_{j=1}^N \bigg(\frac{\mathbb{1}_{d_i \in G_0}}{|G_0|}- \frac{\mathbb{1}_{d_i\in G_1}}{|G_1|}\bigg) \Prob_{i,j} \bv_j = 0 \\
    \Leftrightarrow & \:\:\mathbf{f}^T P\bv = 0 \tag{with $\mathbf{f}_i = \frac{\mathbb{1}_{d_i \in G_0}}{|G_0|}- \frac{\mathbb{1}_{d_i\in G_1}}{|G_1|}$}
\end{align}
In the last step, we obtain a constraint in the form $\textbf{f}^T\Prob \textbf{g}=h$ which one can plug it into the linear program from Section~\ref{optimization_prob}. We call this a \textit{Demographic Parity Constraint} similar to an analogous constraint in fair supervised learning \cite{calders2009building, zliobaite2015relation}. Similar to that setting, in our case also, such a constraint may lead to a big loss in utility in cases when the two groups are very different in terms of relevance distribution.

\paragraph{\textbf{Experiments.}}
\begin{figure*}
    \centering
    \includegraphics[width=1\textwidth]{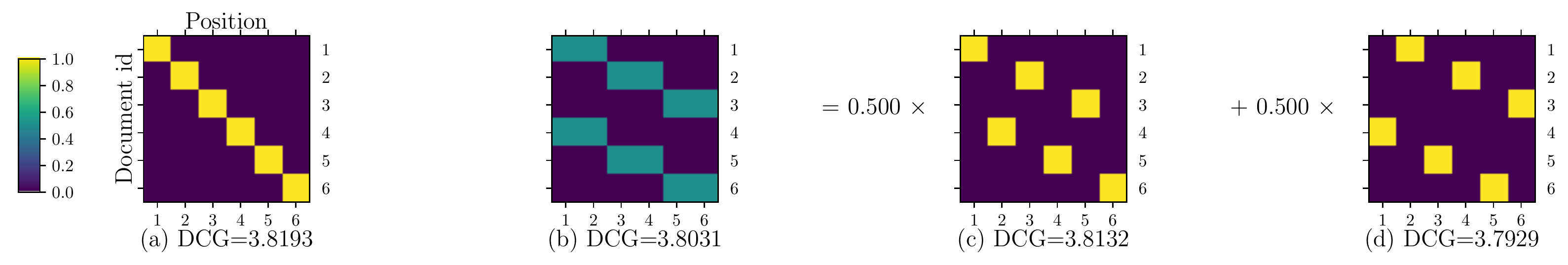}
    \caption{Job seeker example with demographic parity constraint. (a) Optimal unfair ranking that maximizes DCG. (b) Optimal fair ranking under demographic parity. (c) and (d) are the BvN decomposition of the fair ranking.}
    \label{toy:demographic_parity}
\end{figure*}
We solved the linear program in \ref{optimization_prob} twice -- once without and once with the demographic parity constraint from above. For the job seeker example, Figure~\ref{toy:demographic_parity} shows the optimal rankings in terms of $\Prob$ without and with fairness constraint in panels (a) and (b) respectively. Color indicates the probability value. 

Note that the fair ranking according to demographic parity includes a substantial amount of stochasticity. However, panels (c) and (d) show that the fair ranking can be decomposed into a mixture of two deterministic permutation matrices with the associated weights.

Compared to the DCG of the unfair ranking with $3.8193$, the optimal fair ranking has slightly lower utility with a DCG of $3.8031$. However, the drop in utility due to the demographic parity constraint could be substantially larger. For example, if we lowered the relevances for the female group to  $\bu=(0.82, 0.81, 0.80, 0.03, 0.02,\allowbreak 0.01)^T$, we would still get the same fair ranking as the current solution, since this fairness constraint is ignorant of relevance. In this ranking, roughly every second document has low relevance, leading to a large drop in DCG. It is interesting to point out that the effect of demographic parity in ranking is therefore analogous to its effect in supervised learning, where it can also lead to a large drop in classification accuracy \cite{dwork2012fairness}.

\begin{figure}
    \centering
    \includegraphics[width=0.5\textwidth]{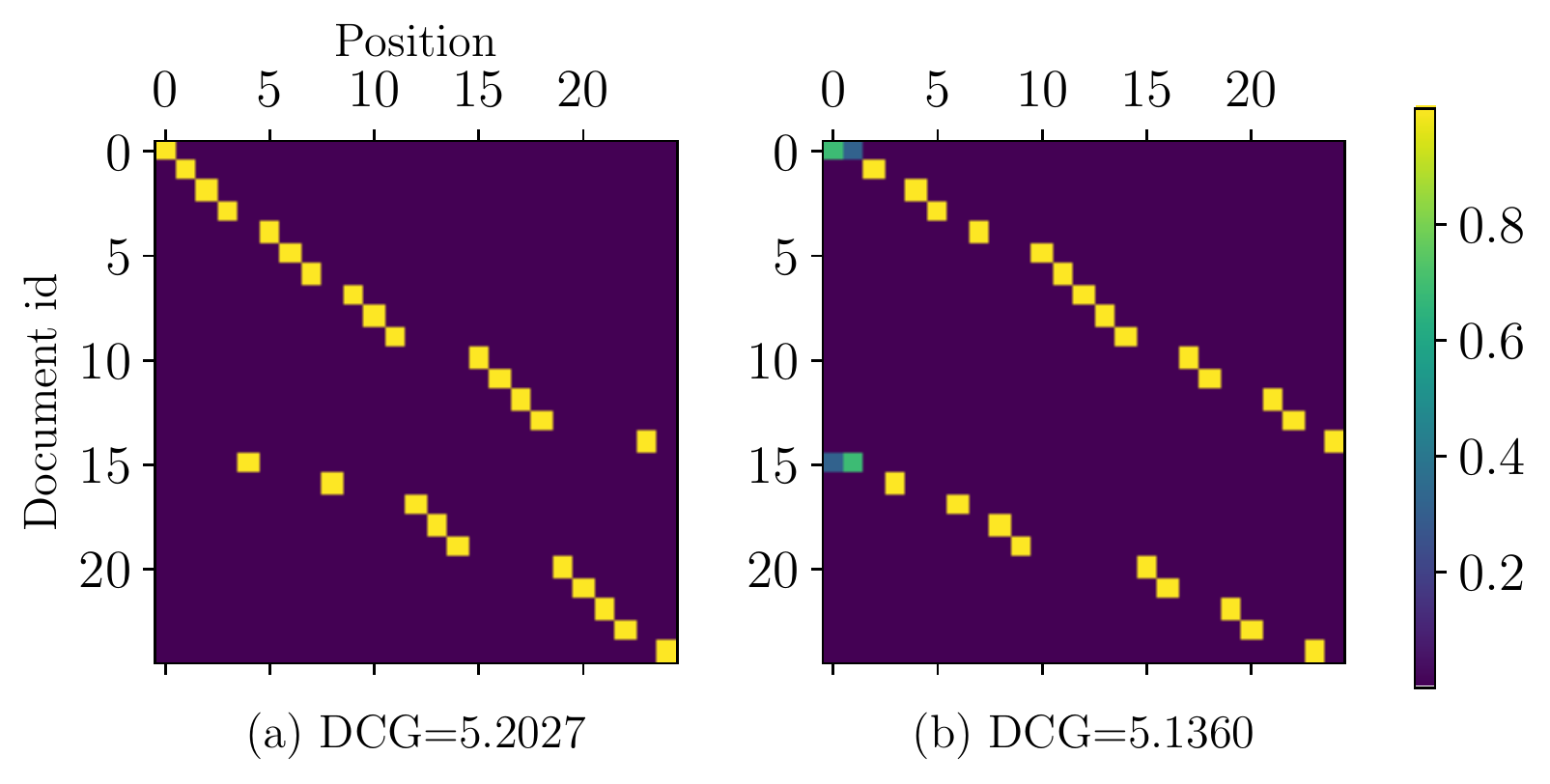}
    \caption{News recommendation dataset with demographic parity constraint. $G_0$: Document id. 0-14, $G_1$: 15-24 (a) Optimal unfair ranking that maximizes DCG. (b) Optimal fair ranking under demographic parity.}
    \label{news:demographic_parity}
\end{figure}
We also conducted the same experiment on the news recommendation dataset. Figure~\ref{news:demographic_parity} shows the optimal ranking matrix and the fair probabilistic ranking along with DCG for each. Note that even though the optimal unfair ranking places documents from $G_1$ starting at position 5, the constraint pushes the ranking of the news items from $G_1$ further up the ranking starting either at rank 1 or rank 2. In this case, the optimal fair ranking happens to be (almost) deterministic except at the beginning.

\subsection{Disparate Treatment Constraints} \label{section:fairness2}

Unlike demographic parity, the constraints we explore in this and the following section depend on the relevance of the items being ranked. In this way, these constraints have the potential to address the concerns for the job-seeker example from the introduction, where a small difference in relevance was magnified into a large difference in exposure. 
Furthermore, we saw that in the image-search example from the introduction that it may be desirable to have exposure be proportional to relevance to achieve some form of unbiased statistical representation. Denoting the average utility of a group with
\[
\ubar{G_k} = \frac{1}{|G_k|}\sum_{\dd_i\in G_k} \bu_i,
\]
this motivates the following type of constraint, which enforces that exposure of the two groups to be proportional to their average utility. 
\begin{align}
    & \:\:\frac{\exposure{G_0}}{\ubar{G_0}} =  \frac{\exposure{G_1}}{\ubar{G_1}} \notag\\
    \Leftrightarrow & \:\:\frac{\frac{1}{|G_0|}\sum_{d_i\in G_0} \sum_{j=1}^N \Prob_{i,j} \bv_j}{\ubar{G_0}}  =  \frac{\frac{1}{|G_1|}\sum_{d_i\in G_1} \sum_{j=1}^N \Prob_{i,j} \bv_j}{\ubar{G_1}} \label{equation:disp-treatment-equality} \\
    \Leftrightarrow & \:\:\sum_{i=1}^N\sum_{j=1}^N \bigg(\frac{\mathbb{1}_{d_i \in G_0}}{|G_0|\ubar{G_0}}- \frac{\mathbb{1}_{d_i \in G_1}}{|G_1|\ubar{G_1}}\bigg)\Prob_{i,j}\bv_j = 0 \label{constraint2}\\
    \Leftrightarrow & \:\:\mathbf{f}^T P\mathbf{v} = 0 \tag{with $\mathbf{f}_i = \bigg(\frac{\mathbb{1}_{d_i \in G_0}}{|G_0|\ubar{G_0}}- \frac{\mathbb{1}_{d_i \in G_1}}{|G_1|\ubar{G_1}}\bigg)$}
\end{align}

We name this constraint a \textit{Disparate Treatment Constraint} because allocating exposure to a group is analogous to treating the two groups of documents. This is motivated in principle by the concept of \textit{Recommendations as Treatments} \cite{schnabel2016recommendations}, where recommending or exposing a document is considered as treatment and the user's click or purchase is considered the effect of the treatment. 

To quantify treatment disparity, we also define a measure called Disparate Treatment Ratio (DTR) to evaluate how unfair a ranking is in this respect i.e. how differently the two groups are treated.
\[\textrm{DTR}(G_0, G_1| \Prob, q) = \frac{\exposure{G_0}/\ubar{G_0}}{\exposure{G_1}/\ubar{G_1}}\]
Note that this ratio equals one if the disparate treatment constraint in Equation~\ref{equation:disp-treatment-equality} is fulfilled. Whether the value is less than 1 or greater than 1 tells which group out of $G_0$ or $G_1$ is disadvantaged in terms of disparate treatment.

\paragraph{\textbf{Experiments.}}

\begin{figure*}[t]
    \centering
    \includegraphics[width=1\textwidth]{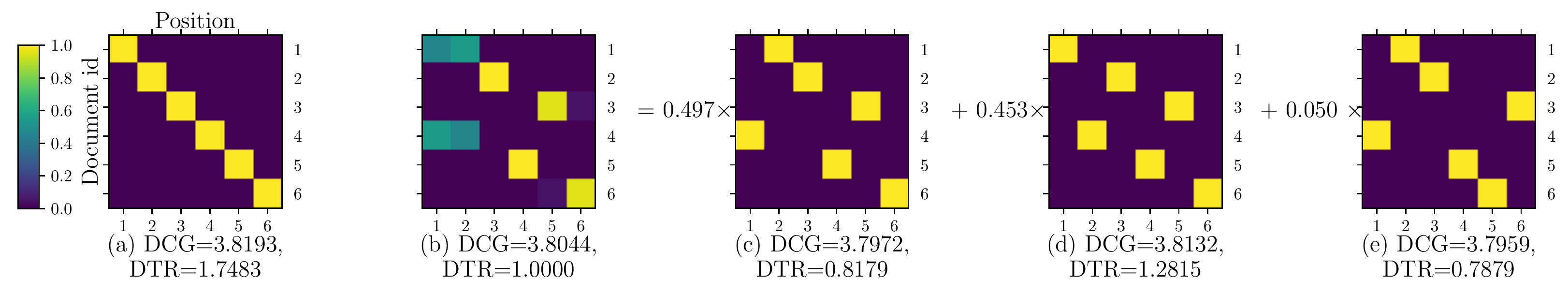}
    \caption{Job seeker example with disparate treatment constraint. (a) Optimal unfair ranking. (b) Fair ranking under disparate treatment constraint. (c), (d), (e) are the BvN decomposition of the fair ranking.}
    \label{toy:disparate_treatment}
\end{figure*} 

We again compute the optimal ranking without fairness constraint, and with the disparate treatment constraint. The results for the job-seeker example are shown in Figure~\ref{toy:disparate_treatment}. The figure also shows the BvN decomposition of the resultant probabilistic ranking into three permutation matrices. As expected, the fair ranking has an optimal DTR while the unfair ranking has a DTR of $1.7483$. Also expected is that the fair ranking has a lower DCG than the optimal deterministic ranking, but that it has higher DCG than the optimal fair ranking under demographic parity. 

\begin{figure}
    \centering
    \includegraphics[width=0.5\textwidth]{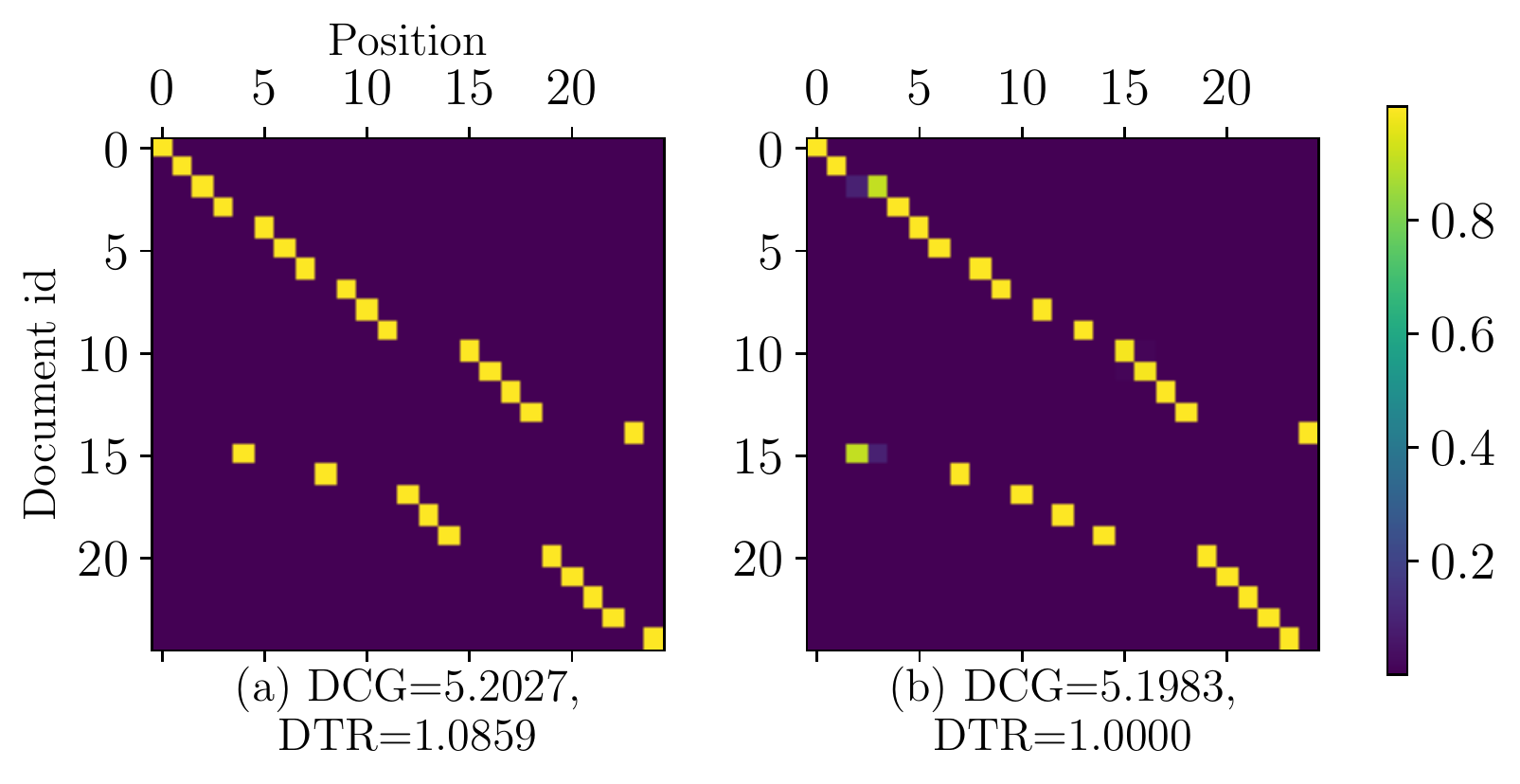}
    \caption{News recommendation dataset with disparate treatment constraint. (a) Optimal unfair ranking. (b) Fair ranking under disparate treatment constraint.}
    \label{news:disparate_treatment}
\end{figure}
We conducted the same experiment for the news recommendation dataset. Figure~\ref{news:disparate_treatment} shows the optimal ranking matrix and the fair probabilistic ranking along with DCG for each. Here, the ranking computed without the fairness constraint happened to be almost fair according to disparate treatment already, and the fairness constraint has very little impact on DCG.

\subsection{Disparate Impact Constraints} \label{section:fairness3}

In the previous section, we constrained the exposures (treatments) for the two groups to be proportional to their average utility. However, we may want to go a step further and define a constraint on the impact, i.e. the expected clickthrough or purchase rate, as this more directly reflects the economic impact of the ranking. In particular, we may want to assure that the clickthrough rates for the groups as determined by the exposure and relevance are proportional to their average utility. To formally define this, let us first model the probability of a document getting clicked according to the following simple click model \cite{Richardson/etal/07}:
\begin{align*}
    P(\textrm{click on document }i) &= P(\textrm{examining } i)\times P(i \textrm{ is relevant})\\
                &= \exposure{d_i} \times P(i \textrm{ is relevant})\\
                &= \bigg( \sum_{j=1}^N \Prob_{i,j}\bv_j \bigg)\times \bu_i
\end{align*} 
We can now compute the average clickthrough rate of documents in a group $G_k$ as
\[
            \ctr{G_k} = \frac{1}{|G_k|}\sum_{i\in G_k} \sum_{j=1}^N \Prob_{i,j} \bu_i \bv_j.
\]
The following \textit{Disparate Impact Constraint} enforces that the expected clickthrough rate of each group is proportional to its average utility:
\begin{align}
    &\:\:\frac{\ctr{G_0}}{\ubar{G_0}}  =  \frac{\ctr{G_1}}{\ubar{G_1}}\\
    \Leftrightarrow &\:\:\frac{\frac{1}{|G_0|}\sum_{i\in G_0} \sum_{j=1}^N \Prob_{i,j} \bu_i \bv_j}{\ubar{G_0}}  =  \frac{\frac{1}{|G_1|}\sum_{i\in G_1} \sum_{j=1}^N \Prob_{i,j} \bu_i \bv_j}{\ubar{G_1}}\\
    \Leftrightarrow & \:\:\sum_{i=1}^N\sum_{j=1}^N \bigg(\frac{\mathbb{1}_{d_i \in G_0}}{|G_0|\ubar{G_0}}- \frac{\mathbb{1}_{d_i \in G_1}}{|G_1|\ubar{G_1}}\bigg)\bu_i \Prob_{i,j} \bv_j = 0\label{constraint3} \\
    \Leftrightarrow & \:\:\mathbf{f}^T P\mathbf{v} = 0 \tag{with $\mathbf{f}_i = \bigg(\frac{\mathbb{1}_{d_i \in G_0}}{|G_0|\ubar{G_0}}- \frac{\mathbb{1}_{d_i \in G_1}}{|G_1|\ubar{G_1}}\bigg)\bu_i$}
\end{align}

Similar to DTR, we can define the following Disparate Impact Ratio (DIR) to measure the extent to which the disparate impact constraint is violated:
\[\textrm{DIR}(G_0, G_1 | \Prob, q) = \frac{\ctr{G_0}/\ubar{G_0}}{\ctr{G_1}/\ubar{G_1}}\]
Note that this ratio equals one if the disparate impact constraint in Equation~\ref{constraint3} is fulfilled. Similar to DTR, whether DIR is less than 1 or greater than 1 tells which group is disadvantaged in terms of disparate impact. 

\begin{figure*}[t!]
    \centering
    \includegraphics[width=1\textwidth]{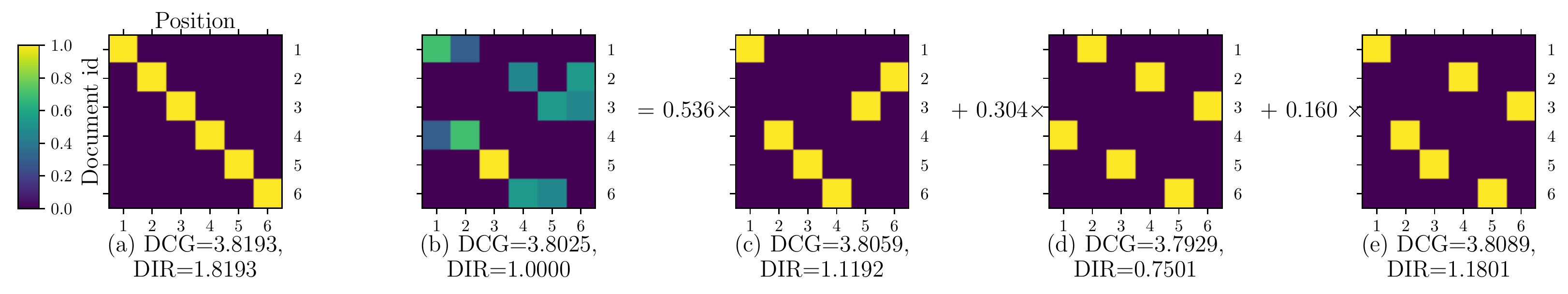}
    \caption{Job seeker example with disparate impact constraint. (a) Optimal unfair ranking. (b) Fair ranking under disparate impact constraint. (c), (d), (e) are the BvN decomposition of the fair ranking.}
    \label{toy:disparate_impact}
\end{figure*}

\paragraph{\textbf{Experiments.}} We again compare the optimal rankings with and without the fairness constraint. The results for the job-seeker example are shown in Figure~\ref{toy:disparate_impact}. Again, the optimal fair ranking has a BvN decomposition into three deterministic rankings, and it has a slightly reduced DCG. However, there is a large improvement in DIR from the fairness constraint, since the PRP ranking has a substantial disparate impact on the two groups.  

\begin{figure}
    \centering
    \includegraphics[width=0.5\textwidth]{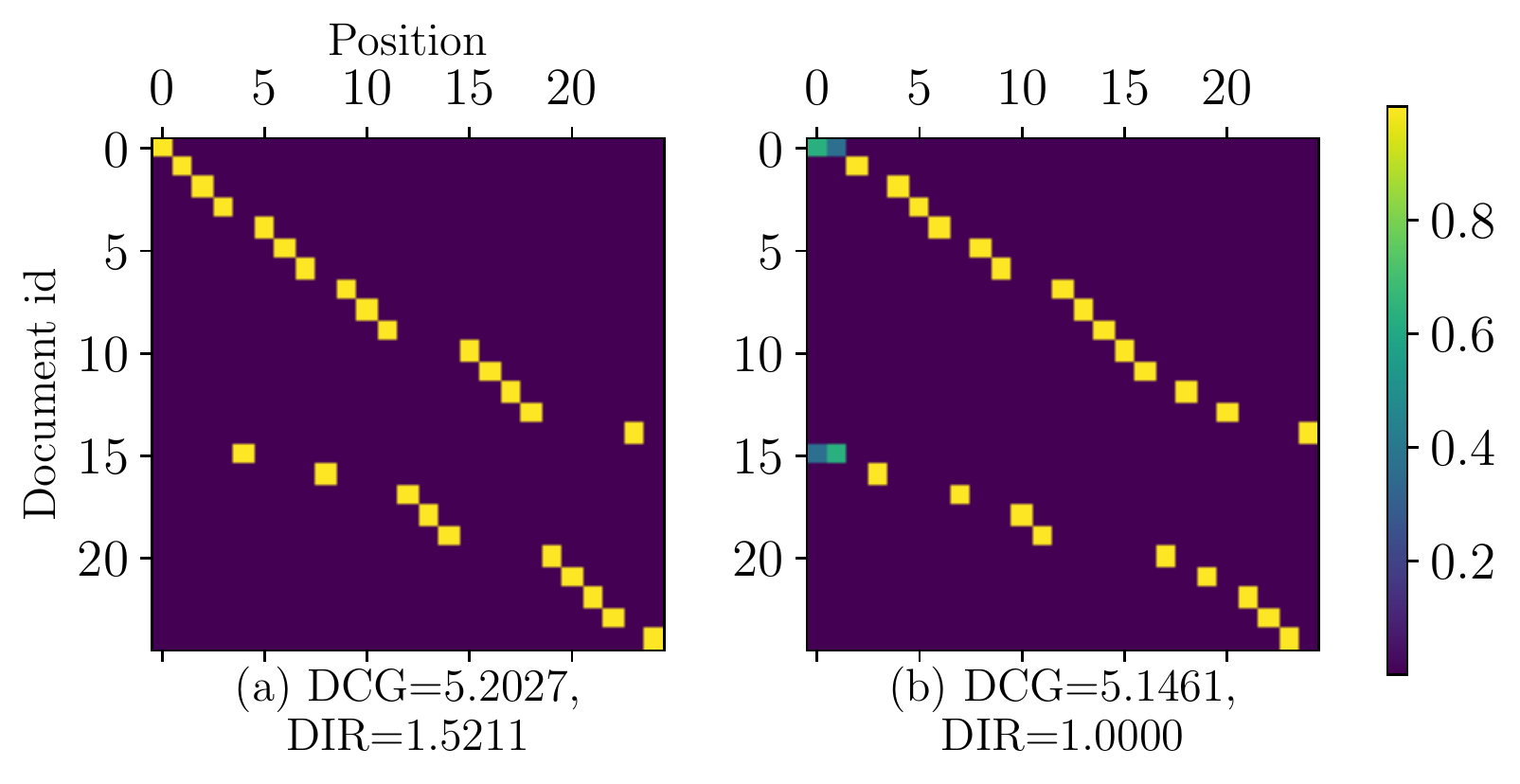}
    \caption{News recommendation dataset with disparate impact constraint. (a) Optimal unfair ranking. (b) Fair ranking under disparate impact constraint.}
    \label{news:disparate_impact}
\end{figure}
The results for the news recommendation dataset are given in Figure~\ref{news:disparate_impact}, where we also see a large improvement in DIR. The DCG is lower than the unconstrained DCG and the DCG with disparate treatment constraint, but higher than the DCG with demographic parity constraint.

\section{Discussion} \label{sec:discussion}

In the last section, we implemented three fairness constraints in our framework, motivated by the concepts of demographic parity, disparate treatment, and disparate impact. The main purpose was to explore the expressiveness of the framework, and we do not argue that these constraints are the only conceivable ones or the correct ones for a given application. In particular, it appears that fairness in rankings is inherently a trade-off between the utility of the users and the rights of the items that are being ranked, and that different applications require making this trade-off in different ways. For example, we may not want to convey strong rights to the books in a library when a user is trying to locate a book, but the situation is different when candidates are being ranked for a job opening. We, therefore, focused on creating a flexible framework that covers a substantial range of fairness constraints.


\paragraph{\textbf{Group fairness vs. individual fairness.}}
In our experiments, we observe that even though the constraints ensure that the rankings have no disparate treatment or disparate impact across groups, individual items within a group might still be considered to suffer from disparate treatment or impact. For example, in the job-seeker experiment for disparate treatment (Figure~\ref{toy:disparate_treatment}), the allocation of exposure to the candidates within group $G_0$ still follows the same exposure drop-off going down the ranking that we considered unfair according to the disparate treatment constraint. As a remedy, one could include additional fairness constraints for other sensitive attributes, like race, disability, and national origin to further refine the desired notion of fairness. In the extreme, our framework allows protected groups of size one, such that it can also express notions of individual fairness. For example, in the case of Disparate Treatment, we could express individual fairness as a set of $N-1$ constraints over $N$ groups of size one, resulting in a notion of fairness similar to \cite{biega2018equity}. However, for the Disparate Impact constraint where the expected clickthrough rates are proportional to the relevances, it is not clear whether individual fairness makes sense, unless we rank items uniformly at random.

\paragraph{\textbf{Using estimated utilities.}} 
In our definitions and experiments, we assumed that we have access to the true expected utilities (i.e. relevances) $u(\dd|\qq)$. In practice, these utilities are typically estimated via machine learning. This learning step is subject to other biases that may, in turn, lead to biased estimates $\hat{u}(\dd|\qq)$. Most importantly, biased estimates may be the result of selection biases in click data, but recent counterfactual learning techniques \cite{joachims2017unbiased} have been shown to permit unbiased learning-to-rank despite biased click data.


\paragraph{\textbf{Cost of fairness.}}
Including the fairness constraints in the optimization problem comes at the cost of effectiveness as measured by DCG and other conventional measures. This loss in utility can be computed as $CoF=\bu^T (\Prob^*-\Prob) \bv$, where $\Prob^*$ is the deterministic optimal ranking, and $\Prob$ represents the fair ranking. We have already discussed for the demographic parity constraint that this cost can be substantial. In particular, for demographic parity it is easy to see that the utility of the fair ranking approaches zero if all relevant documents are in one group, and the size of the other group approaches infinity.

\paragraph{\textbf{Feasibility of fair solutions.}} 
The linear program from Section~\ref{optimization_prob} may not have a solution in extreme conditions, corresponding to cases where no fair solution exists. Consider the disparate treatment constraint
\begin{align*}
    \frac{\exposure{G_0}}{\exposure{G_1}} &= \frac{\ubar{G_0}}{\ubar{G_1}}. 
\end{align*} 
We can adversarially construct an infeasible constraint by choosing the relevance so that the ratio on the RHS lies outside the range that LHS can achieve by varying $\Prob$.
The maximum of the RHS occurs when all the documents of $G_0$ are placed above all the documents of $G_1$, and vice versa for the minimum.
\begin{align*}
    \max\bigg\{ \frac{\exposure{G_0}}{\exposure{G_1}} \bigg\} &= \frac{\sum_{j=1}^{|G_0|}\bv_j}{\sum_{j=|G_0|+1}^{|G_0|+|G_1|} \bv_j}, \tag{all $G_0$ documents in top $|G_0|$ positions}\\
    \min\bigg\{ \frac{\exposure{G_0}}{\exposure{G_1}} \bigg\} &= \frac{\sum_{j=|G_1|+1}^{|G_1|+|G_0|} \bv_j}{\sum_{j=1}^{|G_1|}\bv_j}\tag{all $G_0$ documents in bottom $|G_0|$ positions}
\end{align*}
Hence, a fair ranking according to disparate treatment only exists if the ratio of average utilities lies within the range of possible values for the exposure:
\[ \frac{\sum_{j=|G_1|+1}^{|G_1|+|G_0|} \bv_j}{\sum_{j=1}^{|G_1|}\bv_j} \leq \frac{\ubar{G_0}}{\ubar{G_1}} \leq \frac{\sum_{j=1}^{|G_0|}\bv_j}{\sum_{j=|G_0|+1}^{|G_0|+|G_1|} \bv_j}\]
However, in such a scenario, the constraint can still be satisfied if we introduce more documents belonging to neither group (or the group with more relevant documents). This increases the range of the LHS, and the ranking doesn't have to give undue exposure to one of the groups. 

\section{Conclusions}

In this paper, we considered fairness of rankings through the lens of exposure allocation between groups. Instead of defining a single notion of fairness, we developed a general framework that employs probabilistic rankings and linear programming to compute the utility-maximizing ranking under a whole class of fairness constraints. To verify the expressiveness of this class, we showed how to express fairness constraints motivated by the concepts of demographic parity, disparate treatment, and disparate impact. We conjecture that the appropriate definition of fair exposure depends on the application, which makes this expressiveness desirable.

\begin{acks}
This work was supported by NSF awards IIS-1615706 and IIS-1513692. Any opinions, findings, and conclusions or recommendations expressed in this material are those of the author(s) and do not necessarily reflect the views of the National Science Foundation.
\end{acks}

\bibliographystyle{ACM-Reference-Format}
\balance
\bibliography{main} 

\end{document}